\documentclass{aa}
\usepackage{txfonts}
\usepackage{graphicx}
\usepackage{natbib}
\bibpunct{(}{)}{;}{a}{}{,}

\begin{document}

\title{Evidence of Turbulence-like Universality on the Formation
of Galaxy-sized Dark Matter Haloes
 \thanks{Part of the calculations made use of the NEC SX4/8A 
         supercomputer at the CeNaPAD Ambiental, CPTEC/INPE, Brasil}
}
%\subtitle{}

\author{C. A. Caretta \inst{1, 2}
\and R. R. Rosa \inst{2}
\and H. F. de Campos Velho \inst{2}
\and F. M. Ramos \inst{2}
\and M. Makler \inst{3}}

\institute{
Departamento de Astronom\'ia, Universidad de Guanajuato,
Guanajuato, Gto., M\'exico
\and
Lab. Associado de Computa\c{c}\~ao e Matem\'atica Aplicada,
Instituto Nacional de Pesquisas Espaciais,
S\~ao Jos\'e dos Campos, SP, Brazil
\and
Coordena\c c\~ao de Cosmologia, Relatividade e Astrof\'isica,
Centro Brasileiro de Pesquisas F\'isicas,
Rio de Janeiro, RJ, Brazil
}

\date{Received ... / Accepted ...}

\abstract 
%Context
{Although the theoretical understanding of the nonlinear gravitational
clustering has greatly advanced in the last decades, in particular by
the outstanding improvement on numerical N-body simulations, the 
physics behind this process is not fully elucidated.}
%Aims
{The main goal of this work is the study of the possibility of a 
turbulent-like physical process in the formation of structures,
galaxies and clusters of galaxies, by the action of gravity alone.}
%Methods
{We use simulation data from the Virgo Consortium, in ten redshift
snapshots (from 0 to 10). From this we identify galaxy-sized and 
cluster-sized dark matter haloes, by using a FoF algorithm and
applying a boundedness criteria, and study the gravitational
potential energy spectra.}
%Results
{We find that the galaxy-sized haloes energy spectrum follows closely
a Kolmogorov's power law, similar to the behaviour of dynamically 
turbulent processes in fluids.}
%Conclusions
{This means that the gravitational clustering of dark matter may admit
a turbulent-like representation.}

\titlerunning{Turbulence on Formation of Galaxies}

\maketitle

%--------------------------------------------------------------------------
\section{Introduction}

In a general framework, the standard model of structure formation is 
based on the gravitational Jeans instability criterion \citep[e.g.][]{Lin97}
which neglects viscous, turbulence and diffusion effects.
In this model, the gravity acts to amplify the
tiny random fluctuations in the primordial density field to the
significantly overdense structures we see today (galaxies, groups,
clusters and superclusters).
As long as the density contrast of the fluctuations remains small 
($\delta \ll$ 1) -- in early times for small spatial scales 
(like the ones of galaxies) and until the present time for the largest
scale structures -- the amplitude
of these fluctuations grows linearly. In this regime all scales evolve 
independently \citep[e.g.][]{SC95}, and the random-phase nature of the 
initial density field is preserved. 
Analytically, the equations of motion of the particles
that compose the ``cosmic fluid'' can be successfully solved by means of 
the perturbation theory.
It turns out that, in the hierarchical scenario, fluctuations of increasing
spatial scale successively enter a nonlinear regime, collapse and evolve to
the virial equilibrium. Conversely to the case of the linear regime, 
there is no general exact analytic solution for the evolution of structures
on this stage \citep[e.g.][]{Pad06}.

The search for a theoretical understanding of the nonlinear gravitational 
clustering has greatly advanced in two distinct directions: 
by constructing analytical approximation schemes
and by evolving numerical N-body simulations.
While the attempts on the first approach are limited to the domains of
validity of the approximations, the ones on the last reproduce
quite well most of the observed properties of the actual structures
\citep[see, for instance,][]{Spr05},
but do not clarify completely our 
understanding of the physics behind it.

On the analytical front, the most appealing proposals have been the
ones based on a hydrodynamical approach. The original and simplest 
model of this type is the one called Zel'dovich Approximation 
\citep[ZA,][]{Zel70}, which considers that the velocity of the 
fluid elements can be expressed 
at any time in terms of the initial gravitational
potential, that is, they move in the field without modifying it.
Remarkably this model predicts the 
walls (pancakes), filaments, clumps and voids that characterize the
large scale structure. Nevertheless, the lack of self-gravity leads 
to the dissolution of the structures after the first crossing of the
particles.
Further progress was achieved with a series of approximations 
referred to as adhesion models \citep{Gur89,SZ89,ReF05}.
Mathematically, these models are based on the three-dimensional 
Burgers' equation \citep{Bur40} which 
introduces an {\it ad hoc} artificial viscosity that slows down the 
particles as they approach the pancakes and filaments.
The alternative schemes are mostly extensions of the 
ZA, among which are the more recently proposed Lagrangian 
approximations \citep[see, e.g.,][ and references therein]
{SC95,Mak01,Ber02}.

One possibility that was not much explored is that the formation of
structures is more closely related to hydrodynamics, for example, 
by the existence of a ``turbulent-like'' behaviour of the 
dark matter clustering process.
The idea comes from the characteristic pattern of the large scale 
structure that resembles the
turbulent pattern of fluids, although collisionless dark matter is the 
responsible for such pattern (baryonic matter effects take place 
mostly on smaller scales, while dark energy is expected to be acting 
more recently).

In this paper we use a N-body simulation from the Virgo Consortium to 
to look for signatures of a turbulent-like process in the formation of
structures. For this sake we detect systems of different scales, 
from galactic size subhaloes to supercluster size super-haloes, 
study some of their evolutionary properties, and evaluate the 
gravitational energy of bound systems.

%--------------------------------------------------------------------------
\section{The Virgo N-body Simulation}

Our data came from the intermediate scale cosmological N-body simulations
run by the Virgo Consortium\footnote{available from the web-page \\
{\tt http://www.mpa-garching.mpg.de/Virgo/data\_download.html}},
which are reported in \citet{Jen98}.
Below we briefly describe the main steps and characteristics of these 
simulations \citep[for an extensive review see, for instance,]
[and references therein]{Bert98}.

As usual, the fundamental properties of the N-body simulations of
structure formation in the Universe are defined by the background 
cosmological model and the initial perturbations imposed on this background.
Four versions of the cold dark matter model are available from the
Virgo Consortium project, from which we chose the $\Lambda$CDM one, 
with cosmological parameters 
($\Omega_M$, $\Omega_{\Lambda}$, $h$, $\sigma_8$)\footnote{
$\Omega_M = \bar\rho / \rho_{c} = 8 \pi G \bar\rho / (3 H^{2})$; \\
$\Omega_{\Lambda} = \Lambda c^{2} / (3 H^{2})$; \\
$h = H_0 / 100\,km\,s^{-1}\,Mpc^{-1}$; and \\
$\sigma_8$ is the \textit{rms} mass fluctuation within a top-hat radius 
of $8 h^{-1} Mpc$.}
 = (0.3, 0.7, 0.7, 0.9).
Among the additional (numerical) parameters that characterize the simulation
are the simulation box of side 239.5\,$h^{-1}$~Mpc and the number of 
particles, 256$^3$, with individual masses of 
$6.86 \times 10^{10}\,h^{-1}\,M_{\sun}$.
Comoving spatial coordinates and periodic boundary conditions were
also used \cite[see][for details]{Jen98}.
Ten snapshots are available, at redshifts: 
10.0, 5.0, 3.0, 2.0, 1.5, 1.0, 0.5, 0.3, 0.1 and 0.0.

The initial conditions are set by assigning a position and velocity
to each particle using an algorithm imposing perturbations on an initially
uniform state represented by a ``glass'' distribution of particles
\citep[see, e.g.,][]{Whi96}.
A convenient and efficient method \citep{EDWF} for converting the 
early density fluctuations into these perturbations can be derived from the 
\citet{Zel70} solution: 

\begin{equation}
\vec{x} = \vec{r}/{\mathit a} = \vec{q} + D \ \vec{p} \\
\vec{v} = \frac{d\vec{r}}{dt} - H\vec{r} = a \frac{dD}{dt} 
\vec{p}
\end{equation}

\noindent where $\vec{x}$ is the comoving Eulerian coordinate of a 
particle, $\vec{r}$ is its proper Eulerian coordinate, $a(t)$ is the 
expansion factor (which satisfies the Friedmann equation), $\vec{q}$ 
is the Lagrangian coordinate of the particle denoting its initial 
unperturbed position, $D(t)$ is the growth factor of the linear 
growing mode, $\vec{p}(\vec{q})$ is the displacement field, 
$\vec{v}$ is the peculiar velocity of the particle and, finally,
$H(t) = d\ln a/dt$, is the Hubble parameter. 
The first term of the displacement equation describes the cosmological
expansion, while the second denotes the perturbation. The initial 
velocity field is defined by the initial density fluctuations:

\begin{equation}
\nabla.\vec{p} = - \delta / D
\end{equation}

\noindent where $\delta(\vec{x},t) = [\rho - \bar\rho] / \bar\rho$, \ 
$\rho(\vec{x},t)$ is the local mass density and $\bar\rho(t)$ 
is the spatial mean density.

Since cosmological N-body simulations are intended to follow the 
evolution of the dark matter from the linear regime into the 
deeply nonlinear regime, they must 
begin with the expected distribution of matter, at some early epoch, 
produced by the linear evolution of primordial fluctuations. 
If these fluctuations are Gaussian, they can be fully
described by
the density fluctuation power spectrum (PS), 
$P(k,t) = \langle|\delta_k(t)|^2\rangle$.
This PS can be written as the product of an initial power law, 
produced by the process which generated the fluctuations (typically the 
amplification of quantum fluctuations during inflation), 
a transfer function $T(k)$ representing the linear evolution of each 
mode through the early expansion history (matter domination, 
baryon-photon decoupling, etc.), and the growth factor $D(t)$ governing
the later linear evolution after decoupling:

\begin{equation}
P(k,t) = A k^n T(k)^2 D(t)^2
\end{equation}

\noindent where $A$ is a normalization factor encoding the overall amplitude 
of the initial fluctuactions (generally determined from $\sigma_8$).

For the Virgo simulations, a Harrison-Zel'dovich $n = 1$ spectral index
was used, and also the approximation given by \citet{BE84} for the transfer
function:

\begin{equation}
T(k) = { \{1 + [\mathrm{a}q + (\mathrm{b}q)^{3/2} + 
(\mathrm{c}q)^2]^{\nu}\}^{-1/\nu} }
\end{equation}

%\Delta^2(k) = \frac{A k^4}
%{ \{1 + [aq + (bq)^{3/2} + (cq)^2]^{\nu}\}^{2/\nu} }

\noindent where $q = k/\Gamma$, $\mathrm{a} = 6.4\,h^{-1}$ Mpc, 
$\mathrm{b} = 3\,h^{-1}$ Mpc, $\mathrm{c} = 1.7\,h^{-1}$ Mpc and $\nu = 1.13$.
So, the shape of the PS is determined by the parameter $\Gamma = \Omega_0 h$.
The amplitude $A$, on the other hand, was obtained from the observed 
abundance of galaxy clusters \citep{Jen98}.

Given the initial conditions, that is, the initial position and 
velocity vector of each particle, the simulation is run in about one
thousand timesteps from $z = 30$ to $z = 0$. Newton's law and 
Poisson's equation, in commoving coordinates, are used to evolve 
forward in time the particle trajectories:

\begin{equation}
\vec{x}'' = - \nabla \phi(\vec{x}), \\
\phi(\vec{x}) = - G \sum_{j=1}^{N} \frac{m_j}{ [(\vec{x} - 
\vec{x}_j)^2 + \epsilon^2] }.
\end{equation}

The parameter $\epsilon$ is a softening factor introduced in order to 
prevent the formation of unphysical binaries and large angle particle 
scatterings, and to ensure that the two-body relaxation time is 
sufficiently large to guarantee the collisionless behaviour \citep{EE81}.
The time integration is performed using a second order 
leapfrog scheme \citep{EDWF}.
At each timestep the gravitational force of every particle, 
generated by all others, must be calculated. 
This is a critically time consuming task if done by direct summation, 
so that simplifying parallel algorithms may be used to efficiently compute it.
The Virgo Consortium used a parallel adaptive
P$^3$M algorithm  \citep{CTP95,PC97}, which 
computes long range gravitational forces by smoothing the mass 
distribution in a mesh (particle-mesh, PM), short range gravitational 
forces by direct summation (particle-particle, PP), and applying 
high resolution refinements around strong clustered regions.

%--------------------------------------------------------------------------
\section{Identification of dark matter structures}

There are many methods for detecting dark matter haloes in 
N-body simulations. The most simple, used and tested ones 
are the percolation algorithm 
\citep[or ``Friends-of-Friends'', dubbed FoF,][]{HG82,Ein84,Dav85} 
the ``Spherical Overdensity'' method \citep{LC94}
and the ``DENMAX'' sliding scheme \citep{GB94}.

The FoF algorithm groups together particles that have pair separations smaller 
than a chosen linking length, $\ell$. This linking length is frequently referred
to as $b$ times the mean interparticle separation. The resulting ``groups'' 
are bounded by a surface of approximately constant density:

\begin{equation}
{{n} \over {\bar{n}}} = {{2} \over {(4/3) \pi {\ell}^3}}
{{1} \over {\bar{n}}} = {{3} \over {2 \pi {\ell}^3}}  {\bar{\ell}}^3 
= {{3} \over {2 \pi}} {{1} \over {b^3}} \sim {{1} \over {2 b^3}} 
\end{equation}

Assuming that the density profile of these groups can be approximated by
isothermal spheres, the average density contrast internal to this surface
is given by about 3 times the surface density. 
The larger is $\ell$ (or $b$), the lower is the density contrast and 
the higher is the number of particles linked to the groups. 
In general the value of $b$ is chosen to give a mean overdensity close to 
the value expected for a virialized object in the framework of the spherical
collapse model, $\sim$179. This gives $b=$ 0.2, which seems to be an
appropriate value for detecting cluster-sized haloes 
\citep{LC94,Eke96,Gov99}.
The main advantages of the FoF method are the simplicity (only one free
parameter), the reproduptibility (a unique group catalog for any
chosen value of $\ell$) and the capability of detecting haloes 
of any shape.

The other approach, SO algorithm, identifies density peaks and puts spheres
around them, increasing the radius of the spheres until the average density
contrast reaches a chosen value. It was proposed to overcome the problem 
of FoF that may accidently link two distinct lumps if they are connected 
by a low density bridge of particles. However, it tends to loose the outer
portions of ellipsoidal haloes due to the assumption of spherical symmetry
\citep{LC94}.

The DENMAX algorithm, on the other hand, ``moves'' particles along local 
density gradients toward density maxima, separating halo candidates by 
three-dimensional density valleys. The improvement of this method is 
the application of a self-boundedness check by eliminating particles with 
positive total energy. Unfortunately, this scheme has the drawback 
that the results depend on the level of smoothing of the density 
field \citep{GB94}.

Nevertheless, the three methods above are found to give similar 
results \citep{LC94, Eke96, Aud98}.
The main limitation of these algorithms for detecting structures on N-body
simulations appears for small masses (like the ones of galaxies): they 
do not resolve properly ``sub-haloes'' embedded in host (cluster-sized) 
haloes with the usual parameters.
Improved versions of FoF, SO and DENMAX -- respectively the hierarchical 
FoF \citep[HFoF,][]{Kly99}, the Bound Density Maxima \citep[BDM,][]{Kly99}
and the HOP \citep{EH98} -- and also two step algorithms -- like the 
SUBFIND \citep{Spr01} and the Physically Self-Bound \citep{KP06}
-- have been proposed to better identify sub-haloes in crowded regions.

Here we use a different approach.
It works, in some sense, like the HFoF algorithm.
Since the FoF with $b=$ 0.2 was shown to successfully detect cluster-sized 
haloes, we use this scheme. A cut in a minimum number of particles is used 
to eliminate haloes with masses smaller than the ones expected for 
cluster-sized haloes (see table 1). 
For galaxy-sized haloes (including sub-haloes) we use FoF with a smaller 
linking parameter ($b=$ 0.1). This results in a catalog of haloes smaller 
and denser than the cluster-sized haloes.
Also, the equivalent mean density contrast ($\sim$1\,300) is close to the 
one expected for the Milky-Way if one considers the mass estimate of 
$1.5\times10^{12}\,h^{-1}\,M_{\sun}$
and the radius of $150\,h^{-1}$\,kpc 
\citep[e.g.][]{Smi07, Deh06, Sak03}.
So, this is appropriate for identifying sub-haloes and also for isolated
galaxy-sized haloes.
cD galaxies are the most massive galaxies and, since they are transition 
objects (their halo may be indistinguishable from their parent cluster halo),
they are detected as overmassive galaxy-sized haloes.
They also coincide with their parent cluster position.
For some analyses we excluded these special galaxies by putting a maximum
mass limit.
We also tentatively identify supercluster-sized haloes by relaxing the
linking parameter to $b=$ 1.0 (density contrast about 1.5).
Superclusters are not virialized systems, but structures that probably
have just detached from the Hubble flow.
The volume of the simulation is small for accurate statistics of 
superclusters, but we can select some typical cases for further analysis 
(to be published elsewhere\textbf).

\begin{table}
 \caption{\textsf{Halo identification and selection parameters 
($f$ is the mean density contrast).}}
 \label{Table 1}
  \begin{tabular}{lccrcl}
\hline \hline
Halo size        & $b$ & $\ell$ &  $f$   & \multicolumn{2}{c} {Mass} \\
 \cline{5-6} \\[-0.3 truecm]
                 &     & (Mpc)  &        & ($N_\mathrm{part}$) & ($M_{\sun}$)\\
\hline \\[-0.3 truecm]
Galaxies         & 0.1 &  0.10  & 1\,300 &   2-150   & $> 1\times10^{11}$ \\
Groups/Clusters  & 0.2 &  0.18  &  180   &   $>$ 100 & $> 7\times10^{12}$ \\
Superclusters    & 1.0 &  0.92  &  1.5 & $>$ 15\,000 & $> 1\times10^{15}$ \\
\hline \hline
  \end{tabular}
\end{table}

%--------------------------------------------------------------------------
%\section{Boundedness Criteria}

\begin{figure}
\resizebox{\hsize}{!}{\includegraphics{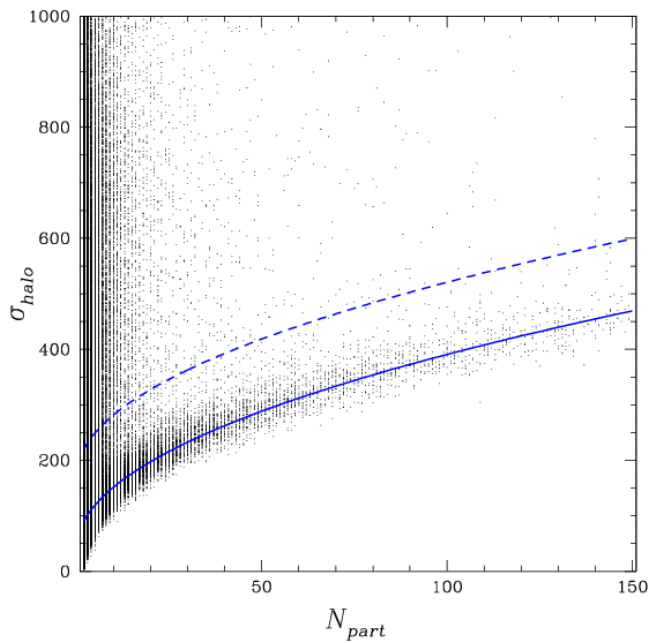}}
\caption{Distribution of galaxy-sized halo masses (N$_{part}$) 
with their respective velocity dispersion ($\sigma_{halo}$),
for redshift 0. 
The solid line is the fit to the points in the main ridge
(a power law $\sigma_{halo} \propto {N_{part}}^{1/2}$ is used),
while the dashed one represents 2.5 rms of dispersion above this fit.}
\label{sigm}
\end{figure}

We also applied a boundedness check for galaxy-sized haloes,  
evaluated in a statistical way.
Figure \ref{sigm} shows the distribution of galaxy-sized halo masses 
(by the number of grouped particles) with their respective velocity 
dispersions.
From the Virial Theorem we expect that, for relaxed bound structures,
these two parameters are correlated, that is, the velocity dispersion
is proportional to the square root of the mass. The ridge of points in
the lower part of the figure follows closely this relation, as can be seen 
by the fit (solid line). So, we considered as bound the haloes below the
dashed line (about 2.5 rms above the fit).
For $z=$ 0.0, 50.7\% of the particles remained isolated after the application 
of the FoF algorithm to search for galaxy-sized haloes, and 10.6\% were 
excluded as unbound haloes. These isolated particles and the particles of
excluded haloes are probably associated to galaxies smaller then our 
detection limit.
A similar behaviour was found for the other redshift snapshots. 

%--------------------------------------------------------------------------
\section{Correlation Functions}

In this section we discuss the clustering properties of the galaxy-sized 
and cluster-sized haloes, found by our percolation scheme, in order to check
the consistency of our object catalogs. We measure the 
two-point (auto)correlation function using the \citet{LS93} estimator:
%Landy \& Szalay (1993)

\begin{equation}
\xi(r) = {{1} \over {RR}} [DD (n_R/n_D)^2 - 2DR (n_R/n_D) + RR]
\end{equation}

\noindent where $DD$, $DR$ and $RR$ are the pair counts, respectively in the 
data-data, data-random and random-random catalogs, and $n_R$ and $n_D$
are the mean number densities in data and random catalogs. We used 
$n_R = n_D$ in all cases. We also applied the 
%Davis \& Peebles (1983) and Hamilton (1993)
\citet{DP83} and \citet{Ham93} 
estimators to our data and found very similar results on
the sampled ranges.

\begin{figure}
\resizebox{\hsize}{!}{\includegraphics{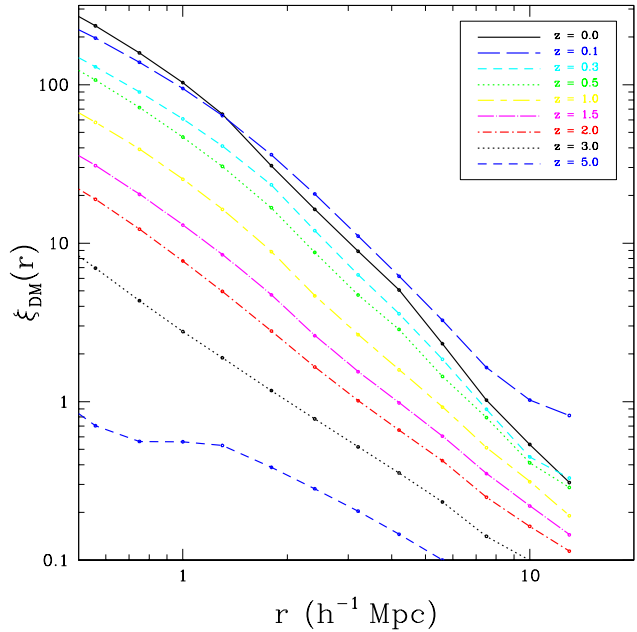}}
\caption{Correlation functions for dark matter particles.}
\label{fcpart}
\end{figure}

Figure \ref{fcpart} shows the two point correlation functions for the 
dark matter particles in the simulation (DMCF), for redshifts 
between 0.0 and 5.0. 
The results for redshifts 0.0, 0.3, 0.5 and 1.0 are very close 
to the ones presented by \citet{Jen98}
on their third top panel of figure 5, as expected.
The $z=$ 0.1 correlation function has a slightly different behaviour on 
scales larger than 1\,$h^{-1}$~Mpc. 
It is a well established fact that the dark matter (mass) correlation 
function has a very complex shape and cannot be represented by a single 
power law in a significant range \citep[e.g.][]{Jen98, Pea99, Ben00, Spr05}. 
Another known property that can
be clearly seen is that the amplitude of the DMCF
evolves considerably with time. Also its slope changes 
systematically with $z$.

\begin{figure}
\resizebox{\hsize}{!}{\includegraphics{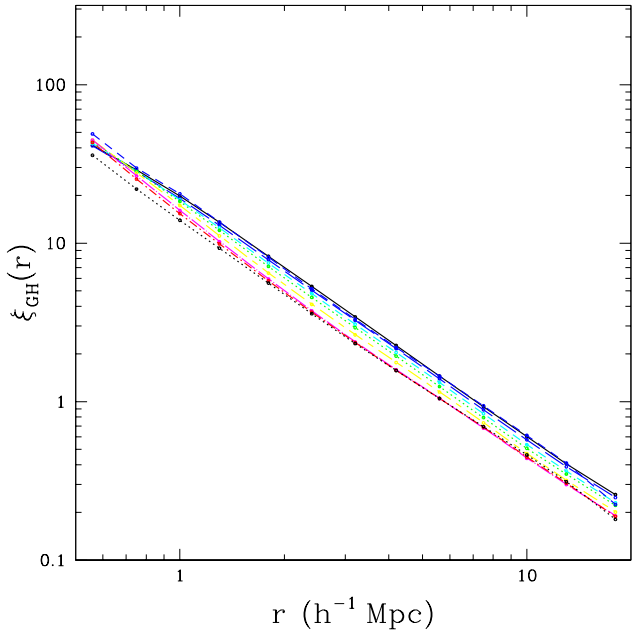}}
\caption{Correlation functions for galaxy-sized haloes.
Line types are the same as on Fig. 2.}
\label{fcgal}
\end{figure}

\begin{figure}
\resizebox{\hsize}{!}{\includegraphics{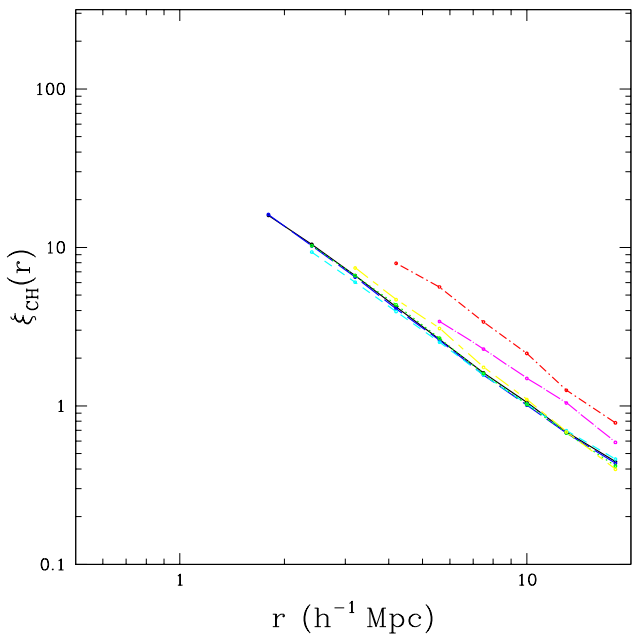}}
\caption{Correlation functions for cluster-sized haloes.
Line types are the same as on Fig. 2.}
\label{fcagl}
\end{figure}

The correlation functions for galaxy-sized haloes (GHCF) are presented on 
Fig. \ref{fcgal}, for the same range of redshifts. Unlike the DMCF,
it shows a shape that resembles closely a power law in all redshifts and 
over a large range of pair separations (here we display the range 
adequately sampled by our data, $0.5 < r < 20\,h^{-1}$~Mpc). 
This behaviour has been 
pointed as really no more than a coincidence \citep[see, e.g.][]{Spr05},
%several processes remarkably conspire to produce this power law.
but, as we shall see in Sect. 5, this might 
be related to the emergence of a turbulent-like 
process in the halo clustering.
On small scales (less than a few Mpc) and low redshifts, the GHCF
has less power than the DMCF,
and the GHCF is said to be ``antibiased''. 
On larger scales, on the other hand, the GHCF
function show similar amplitudes to the DMCF for lower
redshifts, and so it is ``unbiased''.
Furthermore, the GHCF evolves very little
with redshift (its amplitude grows slightly with time).

Figure \ref{fcagl} displays the correlation functions for 
cluster-sized haloes (CHCF). 
One can readily see that, up to redshifts $z \sim 1$, it evolves 
even slower than the GHCF.
At higher redshifts, the CHCF
begins to grow. This is a consequence of the correlation between 
the correlation function of clusters and their masses or richness: 
at higher redshifts, only
the most massive clusters are detected and they are more clustered 
\citep{MW96, Bah04, You05}.

\begin{figure}
\resizebox{\hsize}{!}{\includegraphics{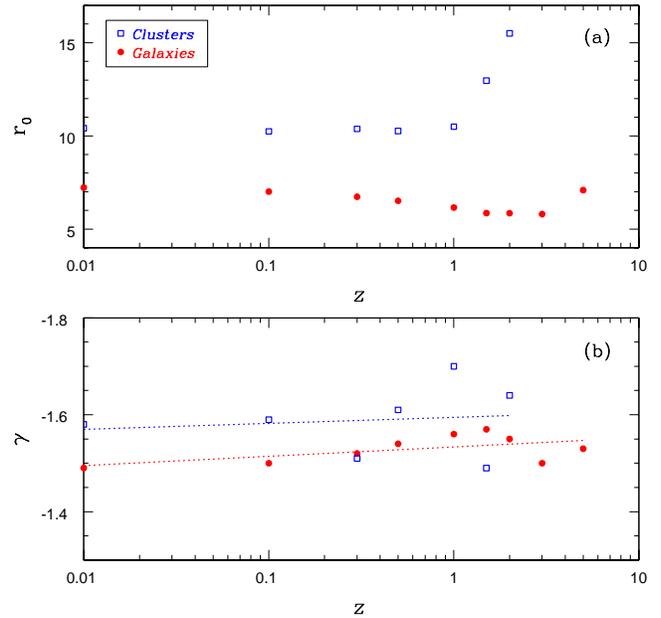}}
\caption{Evolution of the parameters amplitude [panel (a)] and 
slope [panel (b)] for the power law fit to the correlation functions
of galaxy-sized haloes and cluster-sized haloes.
The dotted lines are the linear fits to the data.}
\label{fcres}
\end{figure}

In order to better characterize the evolution of the correlation strength
($r_0$) and slope ($\gamma$) with redshift we fitted power laws to the 
correlation functions of galaxy-sized and cluster-sized haloes 

\begin{equation}
\xi(r) = \left({r \over r_0}\right)^{-{\gamma}}
\end{equation}

\noindent
respectively in the ranges 0.6--18\,$h^{-1}$~Mpc and 2--18\,$h^{-1}$~Mpc. 
The results can be seen on Fig. \ref{fcres}. 
Panel (a) shows the evolution of $r_0$, which grows slightly with time
for both galaxy-sized haloes and cluster-sized haloes (more for the former),
except for the high redshift points for which the biased sampling 
towards more massive objects, as cited above, begins to dominate.
Panel (b) displays the $\gamma$ evolution. There is almost no change in
the slope, except for probable random noise that also grows with redshift
due to the decrease in the number of objects. The mean slope for
galaxy-sized haloes and cluster-sized haloes are, respectively, 1.53
and 1.59, with standard deviations of 0.03 and 0.09.
This values are smaller than the classical 1.8 slope, but consistent 
with recent observational results, especially for galaxies 
\citep[e.g.][]{PB03, Haw03}.

\begin{figure*}
\centering
\includegraphics[width=17cm]{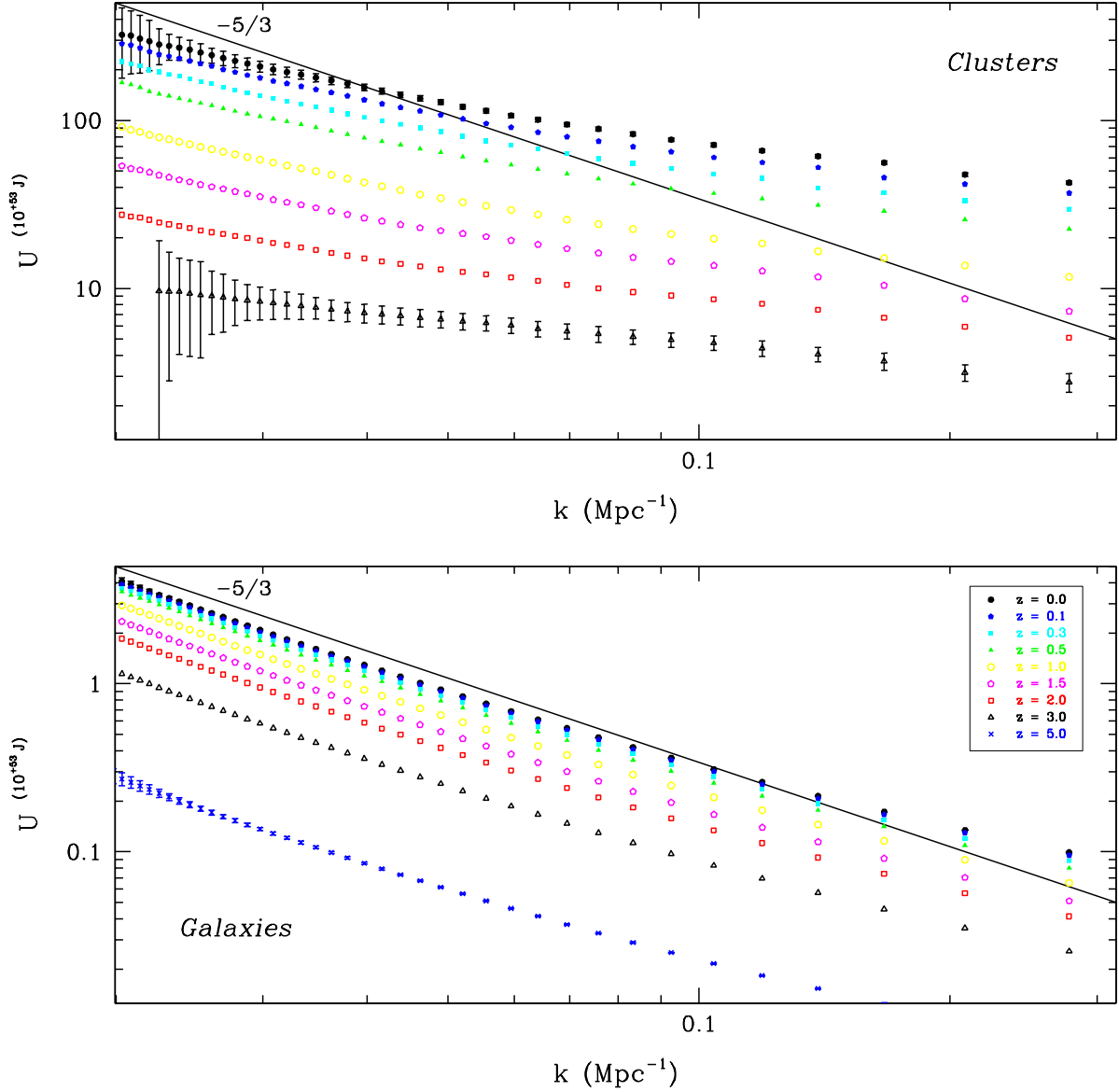}
\caption{Gravitational energy spectra for cluster-sized haloes (above panel)
and galaxy-sized haloes (below panel), for the sampled redshifts.
A reference $-5/3$ power law is plotted as a solid line.}
\label{enspec}
\end{figure*}

%--------------------------------------------------------------------------
\section{Turbulence-like Energy Spectra}

Now we test the hypothesis of the existence of turbulent-like processes 
in the formation of structures, searching for their possible signatures. 
The main characteristic of 3D turbulence is the existence of a kinetic
energy cascade. According to this phenomenology, kinetic energy is
injected in the fluid at large scales by an external mechanical forcing.
Large scale eddies are then deformed and stretched by the fluid
dynamics, breaking into smaller eddies. This process is repeated,
through a hierarchy of smaller and smaller eddies, until the kinetic
energy is finally dissipated into heat by the viscosity of the fluid
\citep{Fri95}. Within a certain range of intermediate scales, the
distribution of energy among turbulence vortices behaves like a 
power law with exponent $-5/3$.
This is the celebrated Kolmogorov's energy spectrum, and
represents the key signature of the ``top-down'' turbulent cascade 
scenario.
Naturally, in an unbounded medium filled with a collisionless set of
particles, the emergence of a turbulent-like behaviour would have a
different physical origin than in the standard hydrodynamic
turbulence. Nevertheless, the effects of particle trajectory crossings
and virialization, as well as the damping effect of the universal
expansion, could lead to similar phenomena as observed in standard
turbulence. For example, the cosmological dynamics of structure
formation could exhibit the power law scaling. This will be the focus
of this section. In other words, we will investigate the energy
distribution of the cosmic structures discussed above (i.e. galaxies
and clusters) to look for the signal of such a turbulent-like behaviour
in the form of a power law scaling.

Since we are not dealing with an usual fluid (dark matter is 
acollisional and, so, not viscous) and we have no energy 
``injection'' in the strict sense, our assumption must be that the 
energy that produces the turbulent-like behaviour is the gravitational
potential energy.
Thus, with the lists of identified galaxy-sized and cluster-sized 
haloes we obtained the energy spectra of the gravitational potential.
This was done by estimating, for each halo and in concentric shells, 
the gravitational energy due to all other haloes:

\begin{equation}
U_{j} = {1 \over 2} G m_j \sum {m_i \over r_{ij}}
\end{equation}

\noindent
where $r_{ij}$ is the distance between the haloes $i$ and $j$.
Then we estimate the mean energy (for all haloes) of each shell
and build the energy spectra with its cumulative value.
The results of this calculation are presented on Fig. \ref{enspec}:
panel (a) for cluster-sized haloes and panel (b) for galaxy-sized
haloes. The stochastic error estimations (${1}/{\sqrt{N}}$)
are plotted only for the first and last redshifts to make the 
viewing of the figure more clear.
The $-5/3$ power law is represented by a solid line.
One can easily notice that the energy spectra for the cluster-sized 
haloes, although following a power law, does not exhibit the 
Kolmogorov's index. 
On the other hand, the one for galaxy-sized haloes follows closely
the $k^{-5/3}$ scaling in the 
wave number range 0.02 to 0.07 (15 to 50 $h^{-1}$~Mpc).

It should be noted that this result differs from the one obtained for 
highly compressible turbulence in molecular clouds \citep{Kri07,Pado07}.
Three-dimensional simulations of supersonic 
Euler turbulence show a velocity power spectrum that gets steeper as the 
Mach number increases, reaching the Burgers slope of $-2$ asymptotically. 
Kolmogorov's $-5/3$ scaling is only recovered by mixing the velocity and 
density statistics, through the computation of density-weighted velocity 
spectra. The very different context of the present paper (namely, 
gravitational clustering of large-scale structures, N-body simulation, 
potential energy spectra) may explain this discrepancy.

To interpret this behaviour we point out that the 
turbulent-like spectra shown in Fig. \ref{enspec} have only the 
inertial regime, without both the typical spectral energy-input 
and dissipation signatures, which are usually found in 
standard turbulent systems.
As originally suggested by \citet{SZ89}, 
% Shandarin \& Zel'dovich 1989,
a simple turbulent scenario 
based on the ZA is compatible with a top-down 
structure formation, where a large structure (pancake), corresponding 
to the integral scale, forms first then fragments into smaller objects 
respecting a cascade process.
However, the observations and simulations have, for many years, favored 
a scenario closer to the hierarchical one.
In such a way, since the gravitational potential energy is purely 
attractive, its variation as a function of the ``wave number'',
defined from the distances among haloes,
could be interpreted as a bottom-up cascade process. 
Physically, because of this attractive nature of gravity, there is 
always an intermittent intrinsic instability that amplifies any 
small homogeneity deviation, as does the equivalent hydrodynamic turbulence. 
Thus, interpreting the results of Fig. \ref{enspec} as the outcome 
of a kind of ``gravitational turbulence'' means that the expansion of 
the universe in the largest scales is driving an inverse fragmentation 
process to compensate its deviation from homogeneity in local scales 
along its evolution.
Furthermore, a turbulent-like behaviour may be a typical and essential 
multi-scaling cosmological feature, requiring, for its robust 
characterization, the use of appropriate measurement techniques 
\citep{Ram02, Wue04, And06}.

%--------------------------------------------------------------------------
\section{Concluding Remarks}

We identified galaxy-sized and cluster-sized haloes in an intermediate 
scale $\Lambda$CDM cosmological N-body simulation by the Virgo Consortium.
For these objects we calculated a gravitational potential energy spectra
and found that the one for galaxy-sized haloes may be well described
by a $-5/3$ power law in the range from 15 to 50 $h^{-1}$~Mpc.
It is interesting to note that the same scaling has been found
in the observational front, for elliptical galaxies, between
potential energy and mass \citep{Mar01}.
We speculate that this scaling of the energy spectrum, shown
in Fig. \ref{enspec}, could be a signature of the Kolmogorov's
universality assumption \citep{Fri95}, where the system dynamical 
structural behaviour is uniquely and universally determined by the 
scaling and its associated mean energy rate. 
From our results, the formation of structures in the nonlinear regime
seems to be driven by a turbulent-like mechanism, associated with 
irregular fluctuations in an unstable gravitational field, 
characterized by combinations of small-scale eddies and larger 
flow-like structures.

It should be noted that a power-law spectrum by itself does not mean 
the existence of turbulence. However, the emergence of a (approximately) 
scale invariant hierarchy of structures as the result of the 
gravitational clustering process, together with the results of 
Fig. \ref{enspec}, 
are consistent with the turbulence-like scenario suggested here.
Also, the space-time patterns seen in the simulated data can be 
interpreted as intermittent nonlinear filaments in a turbulent-like
``gravitational'' fluid \citep[see, for phenomenological comparison 
purposes, the turbulent fluid simulated by][]{She91}.
Because of its importance in our approach, it is worth
mentioning that there is presently no closed theoretical description of 
turbulence \citep[e.g.][]{Sre95, Vel01, Ram01}.
Detailed studies, including analyses of higher resolution cosmological 
simulations and theoretical advances on the models for the nonlinear regime
and turbulence itself, are needed to advance on interpreting the apparent
turbulent-like behaviour of the structure formation processes presented
here.

\begin{acknowledgements}
We thank the Virgo Consortium for providing the Virgo simulation
data, and the CPTEC and CeNaPAD Ambiental for providing the 
NEC SX4/8A supercomputer. The authors would like to thank to 
CNPq, FAPESP and FAPERJ, scientific support agencies. 
C.A.C. acknowledges CNPq grant 382.065/04-2. 
M.M. acknowledges CNPq grants 312425/2006-6 and 486138/2007-0, 
and FAPERJ grant E-26/171.206/2006. 

\end{acknowledgements}

%--------------------------------------------------------------------------
\bibliographystyle{aa}

\end{document}